\documentclass[twocolumn,showpacs,preprintnumbers,amsmath,amssymb,showpacs]{revtex4}

\usepackage{graphicx}
\usepackage{dcolumn}
\usepackage{bm}
\usepackage{amssymb}

\newcommand{\bea}{\begin{eqnarray}}
\newcommand{\ena}{\end{eqnarray}}

\begin{document}

\title{New method to constrain the relativistic free-streaming gas in the Universe}

\author{Wen Zhao}
\email{Wen.Zhao@astro.cf.ac.uk}\affiliation{School of Physics and
Astronomy, Cardiff University, Cardiff, CF24 3AA, United Kingdom}
\affiliation{Wales Institute of Mathematical and Computational
Sciences, Swansea, SA2 8PP, United Kingdom} \affiliation{Department of Physics,
Zhejiang University of Technology, Hangzhou, 310014, People's Republic of China}

\author{Yang Zhang}
\affiliation{Key Laboratory of Galactic and Cosmological Research, Center for Astrophysics,
University of Science and Technology of China, Hefei, 230026,
People's Republic of China}

\author{Tianyang Xia}
\affiliation{Key Laboratory of Galactic and Cosmological Research, Center for
Astrophysics, University of Science and Technology of China,
Hefei, 230026, People's Republic of China}



\begin{abstract}

We discuss a method to constrain the fraction density $f$ of the
relativistic gas in the radiation-dominant stage, by their impacts
on a relic gravitational waves and the cosmic microwave background
(CMB) $B$-polarization power spectrum. We find that the uncertainty
of $f$ strongly depends on the noise power spectra of the CMB
experiments and the amplitude of the gravitational waves. Taking
into account of the CMBPol instrumental noises, an uncertainty
$\Delta f=0.046$ is obtained for the model with tensor-to-scalar
ratio $r=0.1$. For an ideal experiment with only the reduced
cosmic lensing as the contamination of
$B$-polarization, $\Delta f=0.008$ is obtained for the model with
$r=0.1$. So the precise observation of the CMB $B$-polarization
provides a great opportunity to study the relativistic components in
the early Universe.

\end{abstract}



\maketitle


\section{Introduction \label{section1}}

Understanding the cosmic components in the Universe is one of the
main tasks for cosmology.
The current observations from
cosmic microwave background, large scale structure, Type Ia supernova, and etc.,
have already indicated
$\sim72\%$ dark energy, $\sim23\%$ dark matter,
$\sim5\%$ baryons and $\sim 0.005\%$ photons
as the main components in the present Universe.
\cite{map5,sdss,snia}.

With the upcoming of the more precise observations,
it becomes possible and necessary to determine other components.
In this Letter, we shall focus on
the determination of the relativistic components in the Universe.
In addition to the photons and the gravitational wave background \cite{rgw},
these components also include the massless (or tiny massive) neutrinos,
the possible scalar field, the Yang-Mills field dark energy in the scaling
stage \cite{sde,ymc,ymc2},
and some unknown massless (or tiny massive) particles, such as the sterile neutrinos \cite{24}.
As known,
a large relativistic component in the Universe
during the big bang nucleosynthesis (BBN) stage
can enhance the expansion rate of the Universe,
leading to a change the primordial abundances of the light elements.
Thereby, one can constrain the total energy density of the
relativistic components during the BBN stage \cite{bbn},
but unable to  distinguish each component,
as the expansion rate is determined by
the total of all the relativistic components.

If a relativistic component behaves as a free-streaming gas of
massless particles at the photon decoupling,
they will also affect the growth of density perturbations,
in addition to the change of the expansion rate.
So by the observation of CMB  spectra,
especially the temperature anisotropy  spectrum
and the matter perturbation,
one can constrain the fraction density $f$ of
the relativistic free-streaming gas
among all the relativistic components \cite{bashinsky,smith,ichikawa}.
However, there are various degeneracies between
$f$ and other cosmological parameters,
which need to be broken for the method to work.

The stochastic gravitational waves backgrounds, generated in the very early Universe due
to the superadiabatic ampliÞcation of zero point quantum fluctuations of the gravitational field \cite{rgw},
provide a much cleaner way to study the evolution of the Universe.
The effect of the neutrino free-streaming gas on the
spectrum of the relic gravitational waves (RGWs) has been examined
in the previous works \cite{weinberg,a1,a2,miao}.
In particular, it has been found that the neutrino free-streaming gas
causes a reduction of the spectral amplitude by $~20\%$ in the
range $(10^{-16}\sim10^{-10})$Hz, and leaves the other portion of
the spectrum almost unchanged \cite{miao}.

This reduced RGWs leave  observable imprints on the
CMB temperature and polarization anisotropies power spectra
\cite{kamionkowski,xia}.
Especially, the $B$-polarization power
spectrum, only generated by RGWs, is reduced by $(20\%\sim35\%)$ when $\ell>200$.
In Ref. \cite{a3}
it is pointed out that the similar effect can also
be generated by other relativistic free-streaming gas.
In this Letter, we introduce a new method to constrain
the fraction energy density $f$
of the relativistic free-streaming gas
by the future CMB $B$-polarization observations.
It will be shown that the value of $\Delta f$, the
uncertainty of $f$ in the radiation-dominant stage,
strongly depends on the value of tensor-to-scalar ratio $r$,
and is limited by the noise power spectra of the CMB experiments.
For the model with $r=0.1$,
CMBPol experiment can give $\Delta f=0.046$.
If considering the ideal case, where only the reduced cosmic lensing
effect on the $B$-polarization is included,
then one has  $\Delta f=0.008$.


\section{Effects of free-streaming gas on RGWs and CMB polarizations \label{section2}}

Incorporating the perturbations to the spatially flat
Friedmann-Lemaitre-Robertson-Walker space-time, the metric is \bea
ds^2=a^{2}(\eta)[-d\eta^2+(\delta_{ij}+h_{ij})dx^idx^j],\ena where
the perturbations of space-time $h_{ij}$ is a $3\times3$ symmetric
matrix. The gravitational wave field is the tensorial portion of
$h_{ij}$, which is transverse-traceless $h_{ij,j}=0$,
$h_{~i}^{i}=0$. Since the relic gravitational waves are very weak,
$|h_{ij}\ll1|$, one needs to just study the linearized field
equation:
\bea
\partial_{\nu}(\sqrt{-g}\partial^{\nu}h_{ij})=-16\pi
G \pi_{ij}.\label{eq_old}
\ena
The relativistic free-streaming gas
gives rise to an anisotropic portion $\pi_{ij}$, which is also
transverse and traceless. By the Fourier decomposition of $h_{ij}$
and $\pi_{ij}$, for each mode $\bf{k}$ and each polarization,
Eq.(\ref{eq_old}) can be put into the form (see for instance
\cite{weinberg})
\bea
\label{freeeq}
\ddot{h}_k+2\frac{\dot{a}}{a}\dot{h}_k+k^2h_k=16\pi
Ga^2\pi_k,
\ena
where the overdot denotes a conformal time
derivative $d/d\eta$. This equation can be modified to the
following integro-differential equation \cite{weinberg}
\bea
\ddot{h}_k&+&2\frac{\dot{a}}{a}\dot{h}_k+k^2h_k \nonumber \\
&=&-24f\left(\frac{\dot{a}}{a}\right)^2
\int_{\eta_{rd}}^{\eta}\dot{h}_k(\eta')K(k(\eta-\eta'))d\eta',\label{eq_middle}
\ena
where the kernel function in Eq.(\ref{eq_middle}) is
\[
K(x)\equiv-\frac{\sin x}{x^3}-\frac{3\cos x}{x^4}+\frac{3\sin
x}{x^5},
\]
$f\equiv \rho_f/\rho_0$ is the fractional density of
the relativistic free-streaming gas in the radiation-dominant
stage, and  $\eta_{rd}$ is the decoupling time of the relativistic
free-streaming gas.
One has $f=0.41$ for the decoupled neutrino
background with the number of species $N_{\nu}=3$ as the
relativistic free-streaming gas.
However, if the
other relativistic free-streaming gases also exist in the early
Universe,
the value of $f$ should be larger than $0.41$.
On the other hand,
if the neutrinos do not free-stream, due to some possible
couplings \cite{coupling}, then the value of $f$ should be smaller
than $0.41$. So the determination of $f$ provides a chance to
study the relativistic components in the early Universe.

In the analytic approach, Eq.(\ref{eq_middle}) is
approximately reduced to the following form \cite{xia}:
\bea
\ddot{h}_k+2\frac{\dot{a}}{a}\dot{h}_k
+\left[k^2-24f(1-K(0))\left(\frac{\dot{a}}{a}\right)^2\right]h_k=0.
\label{eq_last}
\ena
When $f=0$, this equation returns
to the evolution equation of gravitational waves in the vacuum,
$\ddot{h}_k+2\frac{\dot{a}}{a}\dot{h}_k+k^2h_k=0$ \cite{s1,s2,s3},
which only depends on the evolution of the scale factor $a(\eta)$.
Eq.(\ref{eq_last}) has been solved by perturbations, yielding the
full analytic solution $h_k(\eta)$, from the inflation up to the
present accelerating stage \cite{miao,xia}, and it has been found
that the relativistic free-streaming gas causes a damping of $h_k$
by $\sim 20\%$ in the frequency range
$\nu\simeq(10^{-16},10^{-10})$Hz.

\begin{figure}[t]
\centerline{\includegraphics[width=10cm,height=7cm]{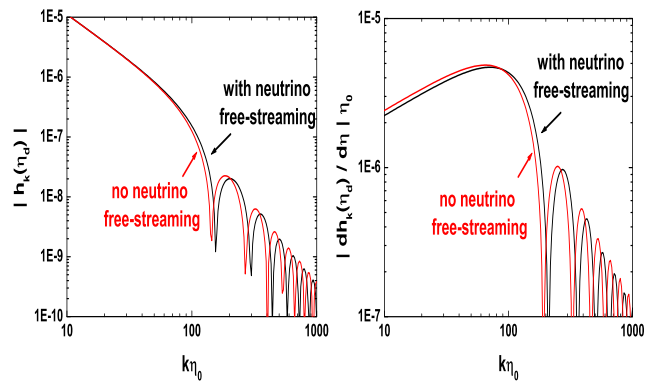}}
\caption{The RGWs $|h_k(\eta_d)|$ and $|\dot{h}_k(\eta_d)|$ at the
decoupling, where we have adopted the parameters of the primordial
power spectra $r=0.1$ and $n_t=0$.}\label{figure1}
\end{figure}

The RGWs can generate the CMB temperature and polarization
anisotropies power spectra $C_{\ell}^{XX'} (XX'=TT,TE,EE,BB)$, by the Sachs-Wolfe
effect
\cite{polnarev,grishchuk,zaldarriaga,kamionkowski2,kamionkowski,ana1,deepak,w2,keating,zhang,xia}.
As shown in Ref \cite{xia}, the mode functions $h_k(\eta_d)$ and
$\dot{h}_k(\eta_d)$ at the photon decoupling time $\eta_d$, i.e.
$z\sim1100$, appear in the integral expressions of the spectra of
CMB temperature and polarization anisotropies.
In
Fig.\ref{figure1},
we plot the quantities $|h_k(\eta_d)|$ and
$|\dot{h}_k(\eta_d)|$ as a function of $k\eta_0$,
where $\eta_0$ is the present
conformal time. The conformal wavenumber $k$ is related to the frequency by $\nu=k/2\pi$, by setting the present scale factor $a(\eta_0)=1$.
We find that the neutrino free-streaming
shifts the peaks of $h_k(\eta_d)$ and $\dot{h}_k(\eta_d)$ to the
right side.
In addition, when $k\eta_0>200$, the amplitudes of
$h_k(\eta_d)$ and $\dot{h}_k(\eta_d)$ are obviously reduced by
$\sim 20\%$, due to the existence of the neutrino free-streaming.

\begin{figure}[t]
\centerline{\includegraphics[width=10cm,height=7cm]{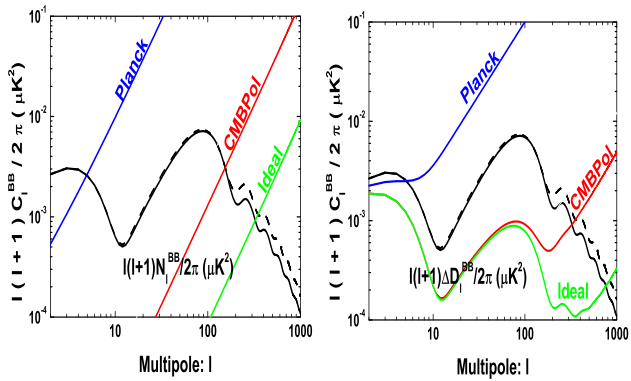}}
\caption{In both panels, the black solid lines denote the CMB $BB$
power spectrum with $f=0.41$ and $r=0.1$, and the black dashed
lines denote the $BB$ spectrum with $f=0$ and $r=0.1$. In the left
panel, we also plot the noise power spectra $N_{\ell}^{BB}$, and
in the right panel, we also plot the quantity $\Delta
D_{\ell}^{BB}$.}\label{figure2}
\end{figure}

The modifications on $h_k(\eta_d)$ and $\dot{h}_k(\eta_d)$ by this
relativistic free-streaming gas leave observable imprints in the
spectra of CMB.
To demonstrate this, the spectra $C_{\ell}^{BB}$
with and without neutrino free-streaming gas are plotted in
Fig.\ref{figure2}.
The $\ell<200$ portion of the spectra is  not
affected much by neutrino free-streaming gas.
Only on the scales
of $\ell>200$, the spectra are modified effectively, i.e. the
reduction of amplitude of $C_{\ell}^{BB}$ by neutrino
free-streaming gas is noticeable only starting from the second
peak. Given the current precision level of observations on CMB,
these small modifications caused by neutrino free-streaming gas
are difficult to detect.
However, as will be shown in the next section,
this modification is expected to be detected by
the future CMB experiments, such as the CMBPol project
\cite{cmbpol}, which are sensitive for the CMB polarization
observations.

\section{Constraint on the relativistic free-streaming gas\label{section3}}

As mentioned, in addition to the decoupled neutrino, there may be
other relativistic free-streaming gases in the early Universe,
which may also modify the RGWs and CMB power spectra.
So by the
observations of the CMB power spectra, especially the
$B$-polarization power spectrum (which is only generated by RGWs),
we can constrain the fraction energy density of all the
relativistic free-streaming gases, which is helpful to understand
the various components in the Universe.

If all the CMB fields are Gaussian random, the power and cross
spectra of the CMB temperature and polarization anisotropies
quantify all the information contained in the observation
\cite{zbg}.
We can use the Fisher information matrix techniques to compare and
contrast the precision,
 to which various surveys can
determine the parameters underlying the power spectra.

The Fisher matrix is a measure of the curvature of the likelihood
function around its maximum in a space spanned by the parameters,
such that the statistical error on a given parameter $p_{i}$ is:
$\Delta p_{i}\simeq ({\bf F}^{-1})_{ii}^{1/2}$
\cite{fisher,fisher2}.
Here we consider the simplest
case, only the fraction density of the
relativistic free-streaming gas, $f$,
is taken as the free
parameter,
and only the CMB $B$-polarization power spectrum is
employed to constrain $f$.
The other cosmological
parameters are assumed to be well determined by the CMB power spectra
$C_{\ell}^{TT}$, $C_{\ell}^{TE}$ and $C_{\ell}^{EE}$ by the future
CMB observations,
so they will be fixed as their fiducial choices
in the data analysis.
Thus the Fisher matrix
$\Delta p_{i}=1/\sqrt{{\bf F}_{ii}}$ for $p_{i}\equiv f$
can be written as \cite{fisher}
\bea \Delta
f=\left[\sum_{\ell}\left(\frac{\partial C_{\ell}^{BB}}{\partial
f}\frac{1}{\Delta D_{\ell}^{BB}}\right)^2\right]^{-1/2}.
\label{fisher}
\ena
Here $\Delta D_{\ell}^{BB}$ is the standard
deviation of the estimator $D_{\ell}^{BB}$ \cite{zbg},
which is
calculable by
\bea \Delta
D_{\ell}^{BB}=\sqrt{\frac{2}{(2\ell+1)f_{\rm
sky}}}(C_{\ell}^{BB}+N_{\ell}^{BB}),
\ena
where $f_{\rm sky}$ is
the cut sky factor. For a special experiment, the noise power
spectrum is calculated by
\bea
N_{\ell}^{BB}=(\Delta_P)^2\exp\left[\frac{\ell(\ell+1)\theta_{\rm
F}^2}{8\ln 2}\right],
\ena
where $\Delta_P$ is the constant noise
per multipole and $\theta_{\rm F}$ is the full width at a half
maximum beam in radians.
We shall discuss three kinds of
future CMB experiments: the Planck satellite, the planned CMBPol
experiment, and an ideal CMB experiment.
Reference sensitivity for
representative CMB polarization experiments are given in Table
\ref{table1} \cite{planck,cmbpol}.
In the ideal case, we have only
considered the reduced lensed $B$-polarization spectrum as the
contamination of $C_{\ell}^{BB}$, which approximately corresponds
to a noise with $\Delta_P\simeq 0.8\mu$K-arcmin \cite{lensing}.

In Fig.\ref{figure2}, we have plotted the noise power spectra
$N_{\ell}^{BB}$ and the uncertainty $\Delta D_{\ell}^{BB}$
compared with the signal $C_{\ell}^{BB}$
in the model with the ratio $r=0.1$,
where we have taken our fiducial choice of the
cosmological parameters as below: $\Omega_b=0.0456$,
$\Omega_c=0.228$, $\Omega_{\Lambda}=0.726$, $\Omega_{k}=0$,
$h=0.705$, $f=0.41$. The perturbation parameters are adopted as
follows: $A_s=2.445\times10^{-9}$, $n_s=0.96$, $\alpha_s=0$,
$n_t=0$.

Fig.\ref{figure2} shows that the modification of $C_{\ell}^{BB}$
by the relativistic free-streaming gas is noticeable only at
$\ell>200$. Since the amplitude of $C_{\ell}^{BB}$ is very small
in this range, only the very sensitive CMB experiments are
expectable to be able to detect this modification.
Fig.\ref{figure2} also shows that,
Planck mission is only sensitive for the
reionization peak of $C_{\ell}^{BB}$. i.e. $\ell<10$.
So it will be not expected to be able to constrain on the
relativistic free-streaming gas in the Universe.
However, for the
CMBPol experiment, the signal $C_{\ell}^{BB}$ is larger than
$\Delta D_{\ell}^{BB}$ when $\ell<300$,
and a detection of this
modification due to the relativistic free-streaming gas
becomes possible.
By solving Eq. (\ref{fisher}),
 we obtain  $\Delta f=0.046$ for the model with $r=0.1$,
and this uncertainty reduced to $\Delta f=0.008$ for the ideal experiment.

\begin{table}
\caption{Instrumental parameters of the CMB experiments
\label{table1}}
\begin{center}
\label{table1}
\begin{tabular}{|c|c|c|c|}
 \hline &~~Planck~~&~~CMBPol~~&~~Ideal~~\\
 \hline $f_{\rm sky}$& 0.8&0.8 &0.8 \\
 \hline $\theta_{\rm F}$~(arcmin)& 7.1& 5& 2\\
 \hline $\Delta_P$~($\mu$K-arcmin) &81.2& 3.1&0.8 \\
 \hline $\Delta f$  (for $r=0.1$) & ...&0.046 & 0.008\\
  \hline
\end{tabular}
\end{center}
\end{table}

\begin{figure}[t]
\centerline{\includegraphics[width=10cm,height=7cm]{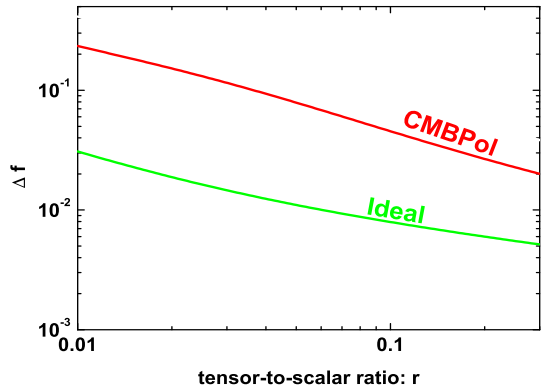}}
\caption{The value of $\Delta f$ depends on the tensor-to-scalar
ratio $r$.}\label{figure3}
\end{figure}

As expected, the value of $\Delta f$ sensitively depends on the
value of tensor-to-scalar ratio $r$.
In Fig.\ref{figure3}, we plot
the value of $\Delta f$ as a function of $r$ for the CMBPol and
ideal experiments.
It is seen that,
with the increasing of $r$,
the value of $\Delta f$ becomes smaller.
For $r=0.3$, one has $\Delta f=0.020$ for CMBPol experiment,
and $\Delta f=0.005$ for ideal experiment.
However,
when $r=0.01$,one has $\Delta f=0.233$ for
CMBPol experiment,
and $\Delta f=0.030$ for the ideal experiment.

\section{Conclusions\label{section4}}

The relativistic free-streaming gas can modify the spectrum of  RGWs
and consequently reduce
the CMB $B$-polarization power spectra at the scale $\ell>200$.
In this Letter, by taking into account the noise power spectra of the
future CMB experiments,
we have presented a constraint on the fraction density  $f$
of the relativistic
free-streaming gas among all the relativistic components during
the radiation-dominant stage.
We find the value of $\Delta f$
strongly depends on the noise of the experiments and the amplitude
of the RGWs.
CMBPol experiment is expected to obtain
$\Delta f=0.046$ for the model with $r=0.1$,
and $\Delta f=0.020$ for the model with $r=0.3$.
For an ideal experiment, where only the
$B$-polarization contamination
by the reduced cosmic lensing effect is included,
we expect to have $\Delta f=0.008$ for the model with
$r=0.1$, and $\Delta f=0.005$ for the model with $r=0.3$.
Our result shows that
the  experiments, like CMBPol,
can provide a great chance to study the relativistic
components in the early Universe.

~

~


{\bf Acknowledgements:}
W.Z. appreciates useful discussions with D.Baskaran and L.P.Grishchuk.
W.Z. is supported by NSFC grants No.10703005 and
No.10775119. Y.Z. is supported by the NSFC No.10773009, SRFDP
and CAS. T.Y.X. is partially supported by Graduate Student
Research Funding from USTC.






\end{document}